\documentclass[iop,apj]{emulateapj}

\usepackage{xspace}

\shorttitle{Mass Dependence of Satellite Quenching}
\shortauthors{Slater \& Bell}

\slugcomment{}

\begin{document}

\title{The Mass Dependence of Dwarf Satellite Galaxy Quenching}

\author{Colin T. Slater and Eric F. Bell}
\affil{Department of Astronomy, University of Michigan,
    500 Church St., Ann Arbor, MI 48109;
    \href{mailto:ctslater@umich.edu}{ctslater@umich.edu},\href{mailto:ericbell@umich.edu}{ericbell@umich.edu}}

\begin{abstract}
We combine observations of the Local Group with data from the NASA-Sloan Atlas
to show the variation in the quenched fraction of satellite galaxies from low
mass dwarf spheroidals and dwarf irregulars to more massive dwarfs similar to
the Magellanic clouds. While almost all of the low mass ($M_\star \lesssim
10^7$ $M_\odot$) dwarfs are quenched, at higher masses the quenched fraction
decreases to approximately 40-50\%. This change in the quenched fraction is
large, and suggests a sudden change in the effectiveness of quenching that
correlates with satellite mass. We combine this observation with models of
satellite infall and ram pressure stripping to show that the low mass satellites
must quench within 1-2 Gyr of pericenter passage to maintain a high quenched
fraction, but that many more massive dwarfs must continue to form stars today
even though they likely fell in to their host $>5$ Gyr ago. We also characterize
how the susceptibility of dwarfs to ram pressure must vary as a function of mass
if it is to account for the change in quenched fractions. Though neither model
predicts the quenching effectiveness {\it a priori}, this modeling illustrates
the physical requirements that the observed quenched fractions place on possible
quenching mechanisms.
\end{abstract}


\section{Introduction}

The shut-off of star formation in galaxies presents one of the most central
features in galaxy evolution but the physical mechanisms at work, along with the
conditions required for quenching, remain poorly constrained. Many mechanisms
have been shown to be capable of shutting off star formation, as the underlying
requirement of denying cold gas to the galaxy can be met in numerous ways.
Broadly speaking, these mechanisms can heat and remove the gas as in ram
pressure stripping \citep{lin83,mayer06} or supernova-driven outflows
\citep{dekel86,ferrara10,sawala10}, or prevent cooling and
accretion of gas onto the galaxy to replenish the gas supply
\citep{efstathiou92,gnedin00,dijkstra04}. In general it can
be easily illustrated that each of these routes for quenching star formation can
plausibly accomplish the task, but it has been difficult to distinguish which of
these mechanisms dominate the quenching process, and under which circumstances.

A fruitful method to help understand the various quenching mechanisms has been
to distinguish between a quenching process that occurs in galaxies which are
satellites of a larger host galaxy and that which occurs in central galaxies
which are the most massive galaxy in their halo
\citep{weinmann06,vandenbosch08,tinker10}. This is motivated both by the
long-standing observation that galaxies in dense environments are preferentially
quenched compared to those in the field \citep{dressler80,postman84,balogh04},
and by the physical differences between mechanisms which could quench satellite
and central galaxies (e.g., an isolated galaxy is unlikely to experience ram
pressure stripping, or satellites are unlikely to merge with each other). This
distinction in mechanisms was readily incorporated into semi-analytic models
\citep{cole94} and tuned to accurately reproduce the distribution of satellite
galaxy colors \citep{font08,weinmann10}. Further observations have sought to
measure the dependence of quenching on both satellite and host halo mass
\citep{wetzel13}. For dwarf galaxies with stellar masses between $10^7$ and
$10^9$ $M_\odot$ the differentiation between satellites and field galaxies is
most acute, as quenched field galaxies are exceedingly rare ($<0.06$\%) in this
range \citep{geha12}. 

The severity of this cut-off in field galaxy quenching provides a strong
motivation to understand how satellite galaxies at similar masses respond to
possible quenching mechanisms. Our primary objective in this work is to
illustrate how the quenched fraction of satellites varies from LMC-mass galaxies
\citep[as in][]{geha12} down to the lowest mass dwarfs we observe in the
Local Group. One of the principal challenges for this is to achieve a
homogeneous selection of galaxies despite the necessarily heterogeneous parent
samples required. Extending our sample to galaxies below roughly $10^8 M_\odot$
in stellar mass requires including satellites of the Local Group, which cannot
be seen elsewhere in wide-area surveys like SDSS. Conversely, galaxies above
this mass are infrequent in the Local Group and a larger survey is required to
obtain meaningful statistics. As a result of these challenges, covering such
large ranges in galaxy mass requires combining heterogeneous samples of the
Local Group dwarfs with larger scale samples like SDSS. This is the strategy we
adopt in this work, which will enable us to illustrate how the quenching
behavior of galaxies changes over five orders of magnitude in mass. From these
measurements, we use N-body simulations to translate the observed quenched
fractions into physical constraints on possible mechanisms, with the intention
of providing guidance to future detailed simulations of the quenching mechanisms
themselves.

In this work we will detail the observed datasets, including the various
corrections for selection effects, in Section~\ref{sec_obs}, and we describe the
resulting quenching fraction behavior in Section~\ref{sec_results}. This result
will then be interpreted in Section~\ref{sec_models} with a comparison to the
distribution of satellite pericenter passage times in Section~\ref{sec_delay}
and separately modeled as ram pressure stripping process in
Section~\ref{sec_rampressure}. We will discuss the implication of these results
and conclude in Section~\ref{sec_conclusions}.

\section{Observations}
\label{sec_obs}

Our goal in this work is to study satellite quenching as homogeneously
as possible over a wide range of mass scales. This necessarily imposes
constraints on our methods. In particular, in the absence of three-space
velocities it is nearly impossible to select only satellite galaxies which are
gravitationally bound to their hosts. We must instead rely on selecting
any galaxies within some representative volume around a host as satellites, in
this case all galaxies within 500 kpc of a host, keeping in mind that some
fraction of these galaxies may be unbound or on first infall onto their host.
All of our comparisons to simulations will be performed with the same selection
process.

Covering a wide range of satellite masses requires us to combine
observations from multiple sources. At the lowest masses we are limited to
galaxies in the Local Group, which is itself a heterogeneous mixture of
individually discovered dwarfs. To put some consistency in this data we use the
compilation of \citet{mcconnachie12}, in which all known galaxies inside of 3
Mpc of the Sun are included. Each galaxy is classified with a ``Morphological''
Hubble type denoting it as either a star forming or a non-star forming type,
though in general this classification is based on studies of resolved stellar
populations rather than morphology alone. The presence of young stars and cold
gas is usually sufficient to identify a Local Group galaxy's star forming
status, but in some cases there is either not sufficient data or a conflicting
set of properties exist, making this determination difficult. These galaxies are
marked as such (e.g., with a Hubble type ``dIrr/dSph'') in the
\citet{mcconnachie12} catalog, and in our figures we include this ambiguity of
classification in the uncertainty on the quenched fraction.  The other main
quantity of interest for this work is the stellar mass of each galaxy, which is
computed from the integrated absolute magnitude assuming a mass-to-light ratio
of 1. This is inherently imprecise, but avoids the much more complex
uncertainties present in dynamical measurements of the total mass of dwarf
galaxies. The uncertainties in mass are relatively small compared to the wide
mass bins we adopt, and thus a change in the overall mass to light ratio or even
a systematic difference between star forming and quenched dwarfs only changes
our reported quenched fractions by a factor smaller than the reported
uncertainty from simply Poisson noise and classification difficulties.

Though the set of known Local Group satellites is certainly incomplete in an
absolute sense, our focus on the relative fraction of star forming versus
quenched galaxies minimizes the impact of this incompleteness. Over the volume
and range of masses we consider here both dSphs and dIrrs are readily detected
in the SDSS, as the red giant branch present in both can be detected out to 
at least 750 kpc \cite[e.g.,][]{slater11,bell11}. The bright stars present in
dIrrs certainly make detection easier, but for any such dIrr that would fall in
our sample a dSph of comparable mass and distance is also likely to be
detectable. That is, over the area covered by the SDSS the detection efficiency
is high for both dSphs and dIrrs, and thus it is unlikely that our measured
quenched fractions are strongly biased by differences in detectability.
Furthermore, as our main result rests on the very high quenched fraction of low
mass satellites, any bias in favor of detecting the brighter dIrrs would only
reinforce this conclusion.

At the mid-range of masses, our sample comes from the NASA-Sloan Atlas of
galaxies \citep[NSA,][]{blanton11}. This sample reprocesses the images from the
SDSS in a manner that better treats the extended surface brightness photometry
required for large galaxies (on the sky) than the standard SDSS pipeline. The
NSA also cross-matches sources with other large surveys and provides stellar
masses estimated with the kcorrect software package \citep{blanton07}.

From this sample of galaxies, we wish to subselect only galaxies that are
satellites of a more massive host. For this we closely follow the method
used in \citet{geha12}, which we summarize here. A sample of candidate
``hosts'' with $M_{K_s} < -23$ (or approximately $2.5 \times 10^{10} M_\odot$ in
stellar mass) was compiled from SDSS and 2MASS and combined with several
different sources of redshift data. This sample is designed to be complete out
to $z = 0.055$, which is the redshift limit of the NSA. Each galaxy in the NSA
was then matched with potential host galaxies by selecting the closest host
galaxy on the sky with a difference in redshift less than 1000 km s$^{-1}$. The
projected distance at the redshift of the host is then recorded as the physical
separation. 

In the work of \citet{geha12} this selection process was used to produce a
very clean sample of isolated field dwarfs. In this work our purpose differs in
that we require a clean sample of satellites with minimal numbers of projected
``interlopers''. This is a much more challenging selection process, since the
significant peculiar velocities of satellites relative to their hosts requires a
wide redshift cut, but such a cut also permits substantial numbers of isolated
galaxies along the line of sight to be included as satellites. This is a
fundamental limitation that cannot be easily remedied by changing the selection
criteria, and instead we attempt to model and correct for the effect. 

We can compute the number of interlopers that fall into our redshift cuts by
constructing mock observations of an N-body simulation. We use the Millennium
simulation for this purpose \citep{springel05}, which simulated a 100
h$^{-1}$ Mpc$^3$ box. This is large enough that the observed volume of the
NSA can fit within the simulation, simplifying the creation of the mock
observations. From the simulation halo catalogs we create a catalog of 
``host'' halos and a catalog of ``dwarfs'', differing only in their halo mass
requirements. We apply the same redshift and projected separation cuts as for
the observed data, then measure the fraction of these selected galaxies that
actually lie within 500 kpc of their host. Since in the NSA our mass cuts are
based on stellar masses, which are not directly available in the N-body
simulation, we convert the stellar mass bins into halo mass bins using the
relation from \citet{moster10}, and also use this to set the limiting mass of a
host halo. 

The measured contamination fraction (interlopers over total number of selected
galaxies) varies smoothly from 65\% at the lowest stellar mass bin in the NSA to
51\% at the highest mass bin. This relatively weak mass dependence limits
the effects of uncertainties in the stellar mass determination, and testing with
an artificially shallow relation (as could be caused by tidal stripping of
satellites) does not substantially affect our results.
The contamination fraction $f_{\rm contam}$ can
directly be used to estimate the corrected quenched fraction $f_Q^\prime$, 
\begin{equation}
f_Q^\prime = f_Q + (f_Q - f_{FQ})\left( \frac{f_{\rm contam}}{1 - f_{\rm contam}}\right),
\end{equation}
where the inclusion of $f_{FQ}$ for the fraction of quenched field galaxies
accounts for the fact that some of the interlopers could themselves be quenched.
Since this factor $f_{FQ}$ is small, the effect of interlopers is to
artificially lower the observed quenched fractions, while the high contamination
fraction causes interlopers to constitute roughly half of the observed
sample. The resulting correction is thus substantial, raising quenched fractions
in the NSA from $\sim 20\%$ to nearly $50\%$ and underlining the importance of
correcting these measurements. We note that interlopers primarily affect
selection of satellite galaxies; the selection of field galaxies like in
\citet{geha12} is much cleaner simply because the broad redshift cut only admits
galaxies to the field sample if they are unambiguously isolated. There is
unfortunately no such unambiguous criteria for satellites.

In addition to the contamination correction, it is also necessary to account for
the relative volumes over which quenched and star-forming galaxies can be
detected in the SDSS. To correct for this we weight each galaxy in the quenched
fraction calculation by the inverse of the volume over which that galaxy could
be detected, which is frequently referred to as a $V_{\rm max}$
correction\footnote{This $V_{\rm max}$ is not to be confused with the maximum
circular velocity of a galaxy, which we will also use in the modeling
section. Sorry.}. This selection bias would otherwise drive the quenched
fractions down, since the brighter star-forming galaxies would be
over-represented. Using the $V_{\rm max}$ correction raises the final quenched
fraction by 15-20\%. We apply this correction only to the NSA sample, as it is
impractical for the Local Group sample where an entirely heterogeneous set of
surveys are responsible for the detection of dwarfs. 

For the NSA sample we distinguish star-forming and quenched galaxies by a
combination of the H$\alpha$ equivalent width (EW) and the $D_n4000$ measure of the
break in the spectrum at $4000{\rm \AA}$. We adopt the criteria of \citet{geha12},
which required quenched galaxies to have an H$\alpha$ EW less than $2{\rm \AA}$ and
required $D_n4000 > 0.6 + 0.1 \log_{10}(M_\star/M_\odot)$. The quenched fraction
is not very sensitive to the specific value of the H$\alpha$ cut; allowing
galaxies with equivalent widths of $4{\rm \AA}$ to be counted as quenched only
changes the resulting quenched fractions by 2-4\%. 

\section{Observational Results}
\label{sec_results}

\begin{figure*}[t]
\epsscale{1.0}
\plotone{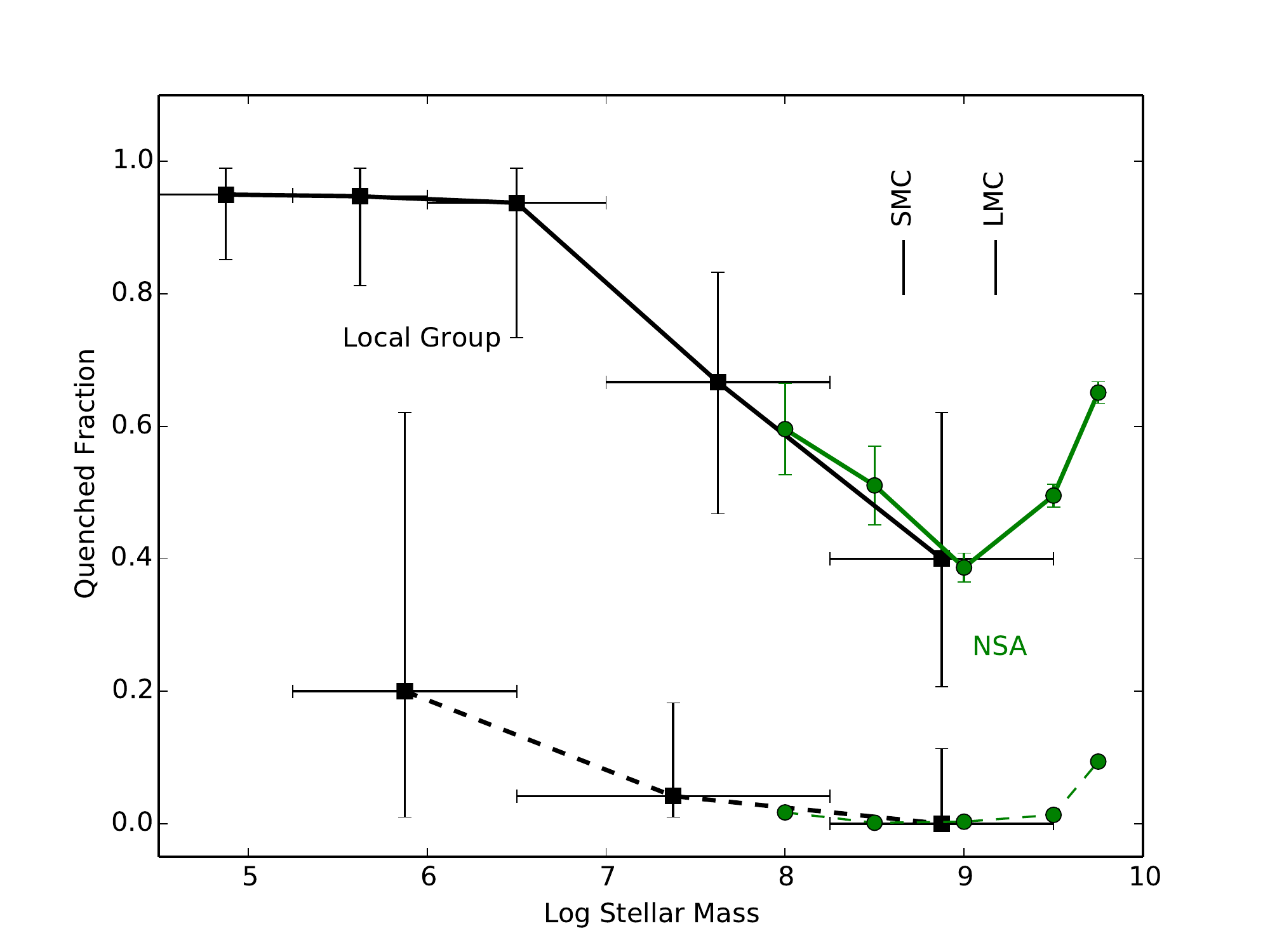}
\caption{Fraction of quenched satellites as a function of galaxy stellar mass
(solid line), along with fraction of quenched field galaxies (dashed line). The
data comprise three samples: dwarfs in the Local Group (black squares), more
massive satellites from the NSA catalog after correction for contamination
(green circles). There is a clear transition near $10^7-10^8$ $M_\odot$ from
nearly ubiquitous quenching of satellites at low mass to much lower quenched
fractions at higher masses.
\label{fig_fracs}}
\end{figure*}

The resulting quenched fractions are shown as a function of satellite mass in
Figure~\ref{fig_fracs}. There is some ambiguity inherent in the classifications
of LG dwarfs into a binary ``star-forming or quenched'' system, so the error
bars on the LG quenched fraction extend from the lowest possible quenched
fraction (assuming all ambiguous galaxies are star forming) to the highest
possible fraction (assuming all ambiguous galaxies are quenched). While this is
clearly not a statistical uncertainty, it does provide an illustration of the
possible range of quenched fractions.

At the lowest-mass end, the data are consistent with nearly all satellites
having no ongoing star formation. The lowest-mass satellite with evidence for recent
star formation is Leo T, with a stellar mass of $1.4 \times 10^5$ $M_\odot$
\citep[under the M/L=1 assumption of][]{mcconnachie12}, but determining the
recent star formation history of such low mass galaxies is challenging
\citep{weisz12}. Other examples of low-mass star forming dwarfs include LGS3
($9.6 \times 10^5$ $M_\odot$) and Phoenix ($7.7 \times 10^5$ $M_\odot$), but
these are substantially outnumbered by quenched dSphs at these masses.
This is in spite of the fact that both types of galaxies down to masses of
$10^5$ $M_\odot$ (e.g., Draco) are well-detected to beyond the limits of the
volume considered here, and that star forming dwarfs are generally easier to
detect.  Not until reaching masses of $10^{6.5} - 10^8$ $M_\odot$ do substantial
numbers of dIrrs begin to reduce the quenched fraction, with galaxies such as
IC 10 ($8.6 \times 10^7$ $M_\odot$), WLM ($4.3 \times 10^7$ $M_\odot$) and IC
1613 ($10^8$ $M_\odot$), for example. It's worth noting that some of these
galaxies may be on initial infall into the LG, and it could be argued that they
are thus not representative of true satellites. While this could of course
modify the absolute quenched fraction depending on the selection criteria, we
argue that this does not affect the mass dependence we seek to illustrate. If
there were no mass dependence in the quenched fraction, then where are the
lower-mass star forming galaxies that are on first infall? Higher mass dwarfs
are not preferentially infalling compared to lower mass dwarfs, as
confirmed with the Via Lactea simulations, and we see little room for selection
effects to cause the mass dependence we observe. The resulting conclusion is
that some changing aspect of the quenching process itself must be responsible
for this effect.

This drop-off in quenched fraction is corroborated by the NSA sample, which
shows similar quenched fractions in the vicinity of 40-60\%. This is
an entirely independent measurement that shares very little in terms of
potential observational biases with the LG data. We have not fine-tuned the
quenching criteria in either sample to create this correspondence, as the
criteria for both samples were originally defined by other works. 
The risk of detection biases related to the host-satellite distance are lessened
in the NSA data, but they are replaced by projection and redshift-related
effects. The principal uncertainty in the NSA measurement is the contamination
correction, which changes the quenched fraction by roughly 20-30\% in each bin.
Even with such a substantial correction, the contamination fraction would have
to be in the range of 80\% or greater to bring the NSA quenched fractions as
high as the seen in the LG.

Low quenched fractions at LMC-range masses are also
seen in the work of \citet{wheeler14}, which reached a similar conclusion with
an alternate methodology. While we have corrected our
observed sample for contamination by redshift-interlopers, \citet{wheeler14} has
created mock observations of their models which include such contamination and
left the observations unchanged. Either process should be equally valid, and the
similarity in resulting values provides an additional confirmation of our
conclusions, but the difference between methods should be noted in making any
direct comparisons. In particular our correction for contamination is necessary
to homogenize the NSA quenched fractions with observations of the Local Group,
which do not suffer from this problem. We also note that the host galaxies of
the NSA satellites are not selected to have a common mass. This may have
implications if the relative mass of satellite and host is an important
determinant of quenching, but in general we expect that the inclusion of
LMC-mass galaxies around much larger hosts than the Milky Way would serve to
raise the quenched fraction rather than lower it, thus minimizing the difference
between the mass ranges rather than artificially increasing the difference. The
conclusion of a substantially lower quenched fraction from $10^{7.5}$ to
$10^{9.5}$ $M_\odot$ appears robust. 

In addition to the fraction of quenched satellites, we also show the fraction of
quenched field galaxies from both the NSA and the Local Group. As
shown by \citet{geha12}, quenched field galaxies are extremely uncommon at
stellar masses below $10^9$ $M_\odot$. The causes of this behavior are beyond
the scope of this work, but we show this to demonstrate that the quenched
fraction of satellites at the masses we are interested in is set primarily by
interactions, and not set by quenching of galaxies in the field. This is
certainly true in the NSA sample, where there is less room for observational
biases to act differentially on field and satellite populations. 

We note that the sample of field galaxies at stellar masses of $10^7$ $M_\odot$ 
may be incomplete, since such intrinsically faint galaxies at
distances of 1 Mpc and greater are observationally challenging. This also
affects field dSphs more than field dIrrs due to their differences in intrinsic
luminosity at fixed stellar mass. For these reasons we do not want to make any
firm statements about the lack of field dSphs. In the LG sample we know of only
a single field galaxy, KKR 25 \citep{makarov12}, that appears quenched, but
it would be difficult to extrapolate from this one galaxy whether a larger
population of field dSphs exists or if this galaxy is somehow peculiar.
In our modeling we will assume that no dwarfs are quenched in the field, but we
acknowledge that this is not yet certain and could be open to revision.

\section{Quenching Models}
\label{sec_models}

Given the changes in the quenched fraction that we see, we would like to
understand how this population-based observation can constrain physical models
for the quenching process. To restate it simply, if we seek to create a scenario
in which 50\% of the high mass dwarfs are quenched, we need to find a criterion
for quenching which is met by only 50\% of the dwarfs at that mass. In this work
we posit two such possible criteria: one which is based on the time since a
galaxy's first pericenter passage around its host, and another based on the
maximum ram pressure experienced by each dwarf. We can then set the parameters
of these criteria such that they reproduce the observed mass dependence of
quenched fraction.

This goal of these models is to illustrate the magnitude of the change in the
quenching criteria with mass required to match the observations, and to put
physical constraints on possible quenching mechanisms based on our observations
of the populations. We note that these simplified models are each taken in
isolation, requiring the change in quenching fraction to be the result of a
single parameter, when in reality there may be several factors that all combine
to produce the observed population. While detailed hydrodynamical simulations
are required for any {\it ab initio} modeling of the quenching process, these
simple models will hopefully demonstrate the magnitude of the problem.

\subsection{Quenching Delay Time}
\label{sec_delay}

\begin{figure}[t]
\epsscale{1.25}
\plotone{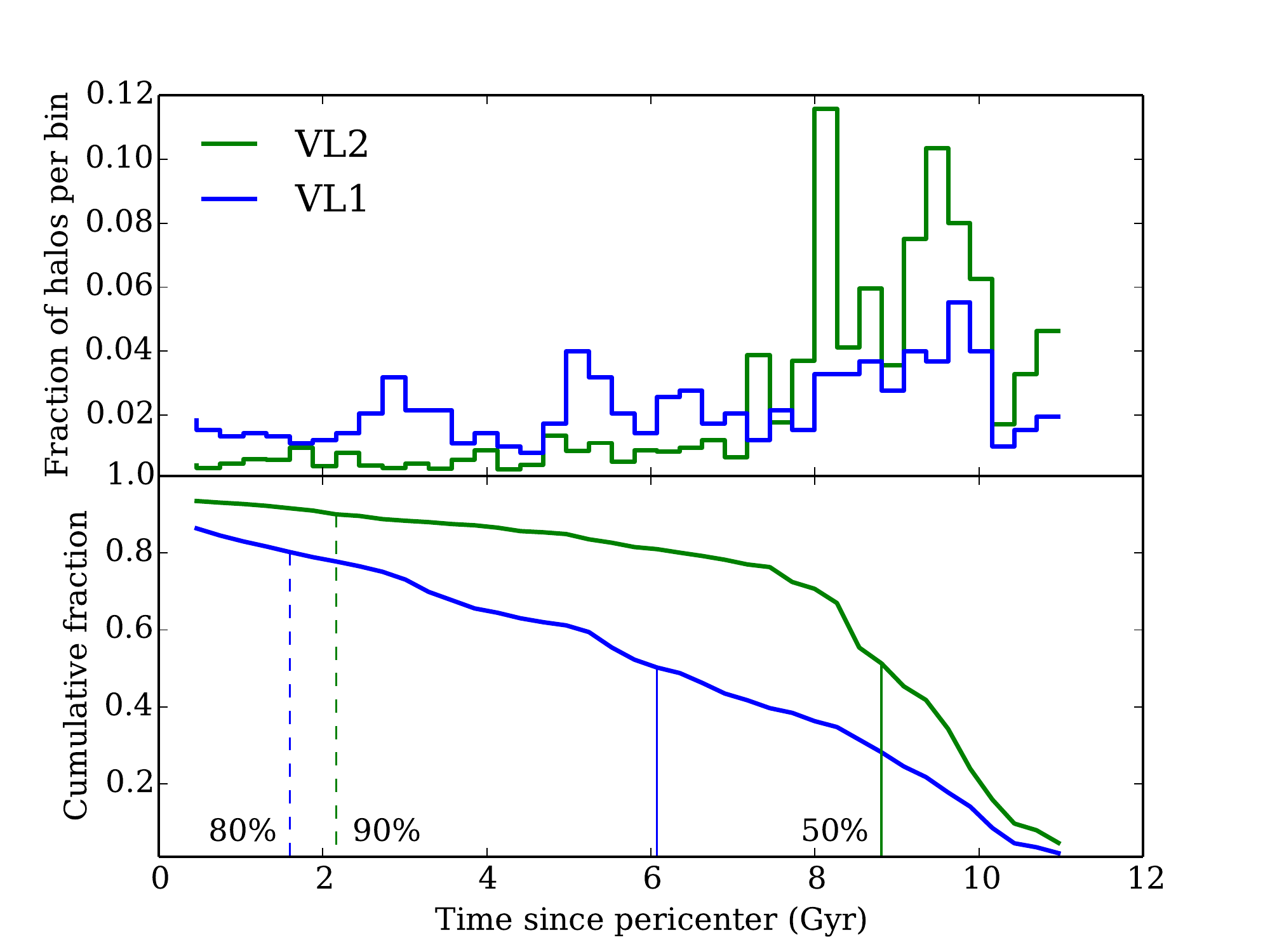}
\caption{Differential (top panel) and cumulative distribution (bottom panel) of
time between a satellite's first pericenter passage and $z=0$, shown for both
Via Lactea simulations. The solid vertical lines indicate where the cumulative
distribution exceeds 50\%, while the dashed vertical lines indicate 80\% and
90\%. This provides a direct estimate of the quenching delay time that would be
required to produce a desired quenched fraction. To reproduce the quenched
fraction of the highest mass LG dwarfs thus requires a delay of 6-9 Gyr between
pericenter passage and quenching, under this model.  To reproduce a quenched
fraction of 80\% or more for low mass galaxies, noting that 10-15\% of selected
halos have not yet experienced a pericenter passage, short quenching times of
order 2 Gyr are required.
\label{fig_peritimes}}
\end{figure}

We first seek to model the changing quenched fraction by positing
that the time since the satellite's first pericenter passage around the host is
the critical parameter. This ``delay time'' model may be interpreted differently
depending on the physical mechanism involved; for example, for large galaxies
falling into clusters the delay time could correspond to a scenario where gas
accretion onto the satellite is stopped upon infall, but some additional time is
required for the star formation to consume the pre-existing gas. This is
primarily of interest when the delay time, as measured in population studies, is
roughly the same duration as the gas consumption timescale for a galaxy. 
Such a delay time has been used to model the quenching of massive galaxies by
\citet{wetzel13}, but we note that our model differs in that we assume
instantaneous quenching after a delay, whereas \citet{wetzel13} have both a
delay and a timescale for star formation to decay. Since we lack both specific
star-formation rates for the dwarfs and sufficient numbers of dwarfs to
disentangle these effects, the assumption of instantaneous quenching will
suffice.

The cumulative distribution of satellite infall times is shown in
Figure~\ref{fig_peritimes}, with the original Via Lactea run in blue and Via
Lactea II shown in green to illustrate the scatter between halo realizations.
From this figure we can see the delay time that would be required for a given
fraction of satellites to remain star-forming in this model. The solid vertical
lines are drawn where the cumulative fraction of satellites that have undergone
pericenter is $0.5$, which is roughly the quenched fraction observed for
satellites at $10^8-10^{10}$ $M_\odot$. The dashed vertical lines are drawn at
cumulative fractions of 80\% and 90\%, which is characteristic of the low mass
quenched fractions observed in the Local Group.

These cumulative pericenter fractions suggest that a rapid quenching process
with a median delay time of $\sim 2$ Gyr is sufficient to reproduce the high
fraction of quenched satellites seen in the LG, though considerable scatter
exists between simulations. This rapid quenching is required to maintain the
high quenched fraction, as recently-infalling satellites would tend to depress
the quenched fraction if they were not quenched quickly. Rapid quenching upon
pericenter also dovetails well with the observed radial distribution of dSphs.
In a previous work \citep{slater13} we showed that reproducing the radial
distribution of quenched LG dwarfs via a close interaction with the host
requires a single such pericenter passage to be sufficient for quenching; any
scenario in which more than one pericenter is required is strongly excluded by
the existence of quenched dwarfs at $\sim 700$ kpc. Rapid removal of gas on a
single pericenter fits both the high quenched fraction and the radial dependence
quite well in the LG.

This short quenching time stands in contrast to the very long gas consumption
timescales of these dwarfs. In general, dIrrs in the field frequently have as
much cold gas as they have stars, if not more \citep{grcevich09}, and at their
mean star formation rates many are unlikely to consume their gas in less than a
Hubble time \citep{hunter85,bothwell09,huang12}. The short timescale for quenching that
we measure reaffirms that the cut-off of gas accretion cannot be responsible for
quenching low mass satellites; such a mechanism would leave far too many star
forming dwarfs in the LG to match the observations. A rapid {\it removal} of
cold gas appears to be necessary to quench a sufficient number of low mass
satellites in a short period of time. 

In comparison to the rapid quenching times for low mass dwarfs,
Figure~\ref{fig_peritimes} suggests that to reproduce the roughly 50\% quenched
fraction at the masses characteristic of the NSA, a median delay time of 6-9 Gyr
is required. This is in line with the conclusions of \citet{wheeler14} that
satellite quenching at these masses is inefficient.  Though this observation is
clear, the cause of such inefficiency is difficult to determine since it can be
the result of a quenching process which is either slow or which operates only on
select dwarfs. It is possible that the time since pericenter is truly a clock
which quenches galaxies that have been satellites for 6-7 Gyr or greater, and
that all unquenched satellites are more recent additions to the LG. This
scenario could arise if infall stopped the accretion of gas and the delay before
a galaxy became quenched was set by the gas consumption time. However, this is
not the only possible interpretation. For example, \citet{wheeler14} suggests
that using the degree of mass loss as a proxy for the strength of interactions
with the host is a more reasonable parameterization for what stops a dwarf's
star formation. While it is possible that time since pericenter is not the
factor that {\it determines} if a galaxy is quenched, it is unavoidable that
some LMC-mass galaxies have been forming stars as satellites for as much as 6-7
Gyr after their first pericenter passage.  If there were not, and only recent
accretions could continue to form stars, the quenched fraction would necessarily
be much higher at these masses. Thus while the evidence is inconclusive as to
whether time is the dominant factor in quenching, whatever does cause quenching
at these masses must permit some satellites to continue to form stars for many
gigayears.

\subsection{Ram Pressure}
\label{sec_rampressure}

\begin{figure}[t]
\epsscale{1.25}
\plotone{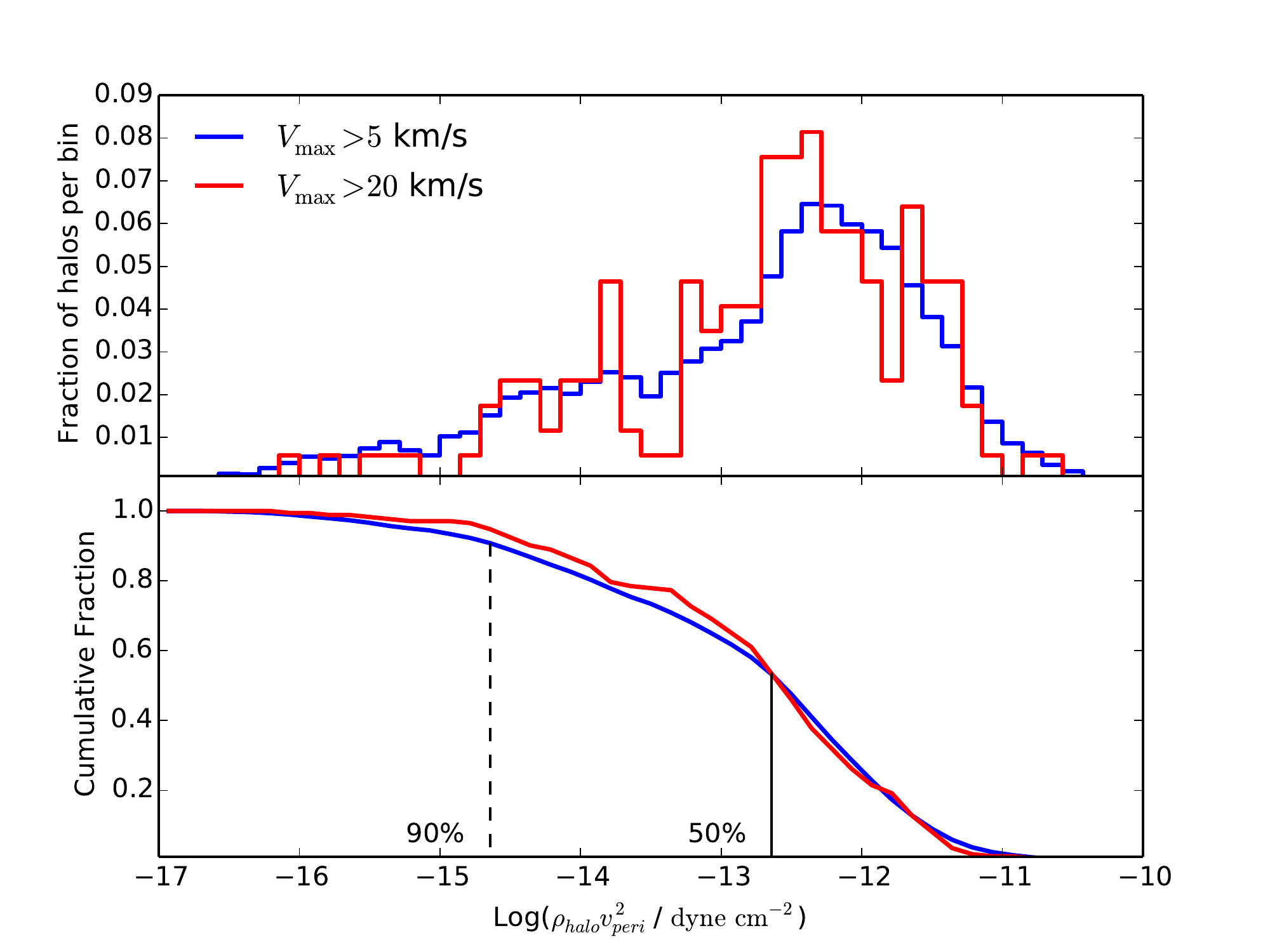}
\caption{Histogram of the maximum ram pressure experienced by halos in the Via
Lactea simulation (top), along with the cumulative distribution of ram pressures
(bottom). The data are split into all halos with $V_{\rm max} > 5$ km/s in blue,
and only halos with $V_{\rm max} > 20$ km/s in red, to show that the ram
pressure distribution is largely independent of satellite mass. From the
cumulative plot we can read the ram pressure required for quenching either 90\%
or 50\% of satellites, and note that the two values differ by roughly a factor
of 100. \label{fig_rampressure}}
\end{figure}

A possible mechanism for the removal of gas from satellites is ram pressure
stripping by hot gas surrounding galaxies and clusters. Initially suggested to
explain the relative infrequence of spiral galaxies in clusters \citep{gunn72},
the presence of hot halos around galaxies has been suggested as a way of
accounting for the deficit of baryons present in the stars and cold gas of
galaxies when compared to the cosmological baryon fraction
\citep{fukugita98,read05,cen99}. Though the observed halos may not be massive
enough to contain all of the missing baryons \citep{benson00,anderson10}, they
may still be able to affect the satellites passing through the hot gas
\citep{lin83,mayer06,mcconnachie07}. 

As noted above, the short timescales required for quenching at low masses
appears to fit naturally with a model where the bulk of the satellite's cold gas
is removed quickly by ram pressure stripping. The ram pressure force
experienced by a galaxy is $\rho v^2$, where $\rho$ is the density of the gas
the satellite moves through and $v$ is its velocity. These two factors are
greatest when a galaxy passes pericenter around its host, thus causing an
``impulsive'' effect on the satellite. The response of a satellite to such a
force will clearly depend on its mass distribution, which determines how
strongly it can hold on to its cold gas. However, the mass distribution of gas,
stars, and dark matter in dIrrs is uncertain, and the magnitude of the restoring
force which resists stripping is difficult to compute {\it a priori} for dwarfs
of differing masses. Our modeling seeks to circumvent this problem by measuring
ram pressure experienced by the population of satellites, and then use this to
constrain how individual galaxies must respond. Other studies in the LG
\citep[e.g.,][]{grcevich09,gatto13} have sought to use the distribution of
stripped and non-stripped dwarfs to constrain the density profile of the Milky
Way's hot halo. We wish to turn this around; using a model of the halo from
X-ray absorption studies \citep{miller13}, what would the quenching criterion
have to be to reproduce the observed quenched fractions?

To compute this, we want to estimate the range of pressures experienced by
satellites as they fall into their host. At each of these
pericenters we can compute the hot gas density, which together with the orbital
velocity gives us the ram pressure force $\rho v^2$.  
To obtain the kinematic information on satellite halos we use both the Via
Lactea and the Via Lactea II simulations \citep{diemand07,diemand08}, which
sought to reproduce a Milky Way-like environment in a dark-matter-only
simulation and together provide a rough estimate of the scatter between halo
realizations. Using these N-body simulations lets us avoid the uncertainties of
hydrodynamic simulations of ram pressure stripping, in which small satellites
are difficult to resolve given the enormous dynamical range required. 
We track the orbit of each surviving subhalo in the simulations
through each of its pericenter passages around either the Milky Way-mass halo or
``Halo 2'', a rough analog of the Andromeda that appears in VL2 \citep[see][for
further details on Halo 2]{slater13,teyssier12}. In finding the pericenter
distances of subhalos we interpolate between snapshots in the simulation, which
prevents pericenter distances from being overestimated due to the limited number
of snapshots. As we showed in \citet{slater13}, we have used the more frequent
snapshots in the VL1 simulation to verify that interpolation does not add
significant errors. 

In this model we assume that the single closest pericenter passage is entirely
responsible for stripping. This is motivated by the strong velocity dependence
of ram pressure, in which pericenter passages should dominate over the rest of
the galaxy's orbit, but also imposed by uncertainties in the cumulative effect
of ram pressure over an extended period of time or multiple pericenter passages.

From the closest pericenter we compute the density of the host galaxy's hot
halo, using the \citet{miller13} model of the Milky Way as an example density
profile. Their work uses a $\beta$-model for the functional form of the profile,
constrained by measurements of X-ray absorption against various extragalactic
and galactic sources, with a total hot gas mass of $3.8 \times 10^{10}$
$M_\odot$ inside of $200$ kpc. We note that this is a measurement of the
present-day halo, and the halo may have been weaker or non-existent 
in the past. In assigning gas pressures seen by halos in the past we should
be using the halo profile present at that time, but the evolution of hot gas
halos is even more uncertain than the structure of halos that exist today. We
thus make as simple of an assumption as is plausible, that the halo has had the
same structure and mass since $z=1$, before which it did not exist. This cut-off
redshift is not critical to the results, and could even be omitted entirely
without significant changes, as the majority of satellites have short enough
orbital periods that they have a pericenter passage after the halo has turned
on. If the density of the hot halo were to change substantially at very late
times then it may have a more significant effect on our results, but any such
halo growth would be entirely an assumption.

The resulting distribution of peak $\rho v^2$ values seen by the subhalos in Via
Lactea is show in Figure~\ref{fig_rampressure}. The top panel shows a histogram
of these values, while the bottom panel shows the cumulative distribution. In
both panels the blue line samples all halos in the simulation with a maximum
circular velocity at $z=0$ of $V_{\rm max} \ge 5$ km/s, while the red line
only includes halos with $V_{\rm max} \ge 20$ km/s. While this division is
arbitrary, we include it to show that there is no significant correlation
between satellite masses and the ram pressures they experience, so we will treat
the results we derive from orbits as essentially independent of mass.

This bottom panel can be read as the fraction of galaxies that have experienced
ram pressure of {\it at least} a given strength; in case we see that 90\% of all
halos have seen ram pressure in excess of $10^{-14.8}$ dyne cm$^{-2}$, while only
50\% have experienced pressures greater than $10^{-12.8}$ dyne cm$^{-2}$. This is
the key result of this model. If we ascribe the entirety of the quenched
fraction change between $M_\star=10^6$ and $10^{7.5}$ $M_\odot$ to changes in a
galaxy's response to a given force of ram pressure, then it is this factor of
100 change in pressure that galaxy models must account for. 

Such a model would need to treat the changing gas densities, stellar disk
densities, and dark matter halo all to obtain a better estimate of the quenching
criterion. This can be seen schematically by rewriting the force
balance from \citet{gunn72} in terms of surface densities \citep{mo10}, 
\begin{equation}
\rho v^2 \sim 2\pi G \Sigma_\star \Sigma_{\rm gas},
\end{equation}
where now the relative distribution of stars and gas may lead to both a
complicated dependence on total mass and could also suggest varying degrees of
partial stripping in some cases. 
Unfortunately these mass distributions are not well constrained observationally,
and the dark matter distribution may also play a role if its contribution to the
restoring force is more significant than the stellar density \citep{abadi99}.
This is difficult to assess from an observational standpoint, as the behavior of
gas which is hypothetically stripped from the disk but remains bound to the
dwarf is unclear.
Even so, better models of dIrrs may not produce more accurate
results if the underlying assumption of a stripping criterion based on force
balance is itself inaccurate. This has been suggested by simulations that better
treat the hydrodynamic instabilities in interactions, resulting in a stripping
that proceeds more via ablation than by impulsive momentum transfer
\citep{weinberg13}. Similarly, the addition of tidal effects during pericenter
passage \citep{mayer06} or internal heating by star formation \citep{nichols11}
could play a significant role in determining a satellite's respsonse to ram
pressure and particularly the dependence on satellite mass. The sum of these
uncertainties both in models of dIrrs and in the physics of stripping limit our
ability to provide a more detailed explanation for the evolution in stripping
efficiency, but the magnitude of the effect is clearly demonstrated in the range
of ram pressure forces experienced.

\section{Discussion and Conclusions}
\label{sec_conclusions}

We have shown that the fraction of quenched satellite galaxies undergoes
significant variation across masses ranging from low mass dwarfs around the
Milky Way and Andromeda to more massive satellites in the LG and beyond. This a
measurement spanning five orders of magnitude in mass, which highlights the
commonality of satellite quenching as a phenomenon but conversely the large span
of masses should also temper our surprise that a complex process like quenching
exhibits varied behavior at different masses. The structure of galaxies across
this range of masses changes substantially, and hence their strongly differing
response to environmental factors may be be a reflection of that fact.

We argue that our conclusion of a rapid quenching process for low mass
satellites is unavoidable given the ubiquity of quenched satellites at these
masses. The speed of quenching immediately places a constraint on plausible
mechanisms, and the rapid removal of gas by ram pressure stripping appears to be
a logical possibility. Quenching processes that proceed on the gas consumption
timescale are difficult to reconcile with the observations. At the masses of the
NSA sample, where the quenched fraction is closer to 50\% than 90\%, the long
delay times leave the question of physical mechanisms open. Here the issue
of a time delay may interact with repeated pericenter passages to remove gas
only gradually in these massive dwarfs. Such a scenario is both beyond the
capabilities of our model and poorly understood physically.

Our model of ram pressure stripping has sought to illustrate how dwarfs of
differing masses must respond to ram pressure, if it is the dominant source of
quenching. This method turns the observed quenched fractions into a value for
the cut-off ram pressure between stripping and leaving a galaxy to continue
forming stars, which provides a characterization of the forces at work. As with
the delay time, it is possible that additional factors add complications to our
picture of a simple ram pressure cut-off. For instance, the inclination of the
disk of a dIrr as it falls into the galaxy may tip the balance if it would
otherwise be on the cusp of being stripped. We argue that our characterization
of the factor of 100 change in ram pressure seen by 50\% and 90\% of satellites
provides an estimate of the average behavior, which may not apply to each
satellite individually. 

While our modeling has attempted to assess the general characteristics of
satellite quenching, our models are clearly not the {\it ab initio} models of
quenching that could explain the mechanisms behind the observed behavior. We
emphasize the importance of attempts to construct such models in order to narrow
what is presently a wide-open range of physical processes that are suggested to
affect quenching. Accurate modeling of the input dIrr galaxies to be stripped is
also critical in this effort, since our understanding of the stripping process
requires detailed knowledge of the systems to be stripped. We have shown at a
basic level what evolution in this effectiveness one might expect with mass, but
this does not attempt to account for the changes in dwarf structure with mass.
What was set up in \citet{gunn72} as a simple force balance between ram pressure
and the restoring force likely has substantial uncertainties on both sides.

We also must emphasize an important caveat of our study, which is that all of
our data below $M_\star = 10^{7.5}$ $M_\odot$ comes from satellites of the Milky
Way and Andromeda. While we argue that these data are robust, the limited number
of systems makes it impossible to know if the high quenched fractions are truly
universal across Milky Way-like systems, or whether they are a peculiar result
tied to the specific accretion history of the Local Group. This is a
particularly important question in the light of results suggesting that the
quenched fraction of satellites is dependent on whether or not the central
galaxy is forming stars, an observation referred to as ``galactic conformity''
\citep{weinmann05,phillips14}. We note that our Local Group results show a high
fraction of quenched satellites around what are clearly star-forming hosts (the
Milky Way and Andromeda). This perhaps illustrates the lower limit at which
galactic conformity is effective; the lowest mass dwarfs appear to quench
regardless of their host. 

Despite this potential complication at LMC-masses, in the absence of any further
information our best estimate of the quenching behavior comes from assuming that
our galaxy is ``average'' and does truly represent a universal behavior, but we
would surely have greater confidence if observations of other systems could
confirm this universality rather than leave it as an assumption. This is but
one of many subjects that stand to gain from the development of larger samples
of dwarfs beyond the Local Group.

\acknowledgments

This work was supported by NSF grant AST 1008342. We again thank the Via Lactea
collaboration for making their simulation outputs publicly available, and M.
Geha for both productive discussions and for providing the host galaxy catalog.

\end{document}